\documentclass[reprint,showkeys,nofootinbib,amsmath,amssymb,aps,prd,]{revtex4-1}
\usepackage[marginal]{footmisc}
\usepackage{graphicx}
\usepackage{dcolumn}
\usepackage{bm}
\usepackage{float}
\usepackage{wrapfig}
\usepackage{subfigure}
\usepackage{amsmath,amssymb}
\usepackage{booktabs}
\usepackage{url}
\usepackage{color}

\newcommand{\be}{\begin{equation}}
	\newcommand{\ee}{\end{equation}}
\newcommand{\ba}{\begin{eqnarray}}
	\newcommand{\ea}{\end{eqnarray}}

\begin{document}

	\title{Constraints on ultra-slow-roll inflation from the third LIGO-Virgo observing run}

	\author{Bo Mu$^{1,2}$}
	\email{mubo22@mails.ucas.ac.cn}
	\author{Gong Cheng$^{1}$}
	\email{chenggong@itp.ac.cn}
	\author{Jing Liu$^{3,4}$}
	\email{liujing@ucas.ac.cn}
	\author{Zong-Kuan Guo$^{1,2,5}$}
	\email{guozk@itp.ac.cn}
	
	\affiliation{$^{1}$CAS Key Laboratory of Theoretical Physics, Institute of Theoretical Physics, Chinese Academy of Sciences, P.O. Box 2735, Beijing 100190, China}
	\affiliation{$^{2}$School of Physical Sciences, University of Chinese Academy of Sciences, No.19A Yuquan Road, Beijing 100049, China}
	\affiliation{$^{3}$International Centre for Theoretical Physics Asia-Pacific, University of Chinese Academy of Sciences, 100190 Beijing, China}
	\affiliation{$^{4}$Taiji Laboratory for Gravitational Wave Universe, University of Chinese Academy of Sciences, 100049 Beijing, China}
	\affiliation{$^{5}$School of Fundamental Physics and Mathematical Sciences, Hangzhou Institute for Advanced Study, University of Chinese Academy of Sciences, Hangzhou 310024, China}

	\begin{abstract}
		The nonattractor evolution in ultra-slow-roll~(USR) inflation results in the amplification of superhorizon curvature perturbations and then induces a strong and detectable stochastic gravitational wave background. In this paper, we search for such a stochastic gravitational wave background in data from the third LIGO-Virgo observing run and place constraints on the USR inflationary models. The $e$-folding number of the USR phase are constrained to be $\Delta N \lesssim 2.9$ at the 95\% confidence level and the power spectrum of curvature perturbations amplified during the USR phase is constrained to be $\log_{10}P_{R\mathrm{p}}<-1.7$ at the scales $2.9\times10^5 ~\mathrm{pc^{-1}} \lesssim k \lesssim 1.7\times10^{11}~\mathrm{pc^{-1}}$. Besides, we forecast the ability of future experiments to constrain USR inflation, and find $P_{R\mathrm{p}}\lesssim 10^{-3.6}$ for LISA and Taiji, $P_{R\mathrm{p}}\lesssim 10^{-3.3}$ for Cosmic Explore and Einstein Telescope, $P_{R\mathrm{p}}\lesssim 10^{-5.5}$ for DECIGO and Big Bang Observer and $P_{R\mathrm{p}}\lesssim 10^{-5.2}$ for Square Kilometre Array.
	\end{abstract}
	
	\maketitle
	\section{Introduction}
	The Advanced LIGO and Advanced Virgo provide a novel and prospective method for observing the Universe and testing gravity theories by detecting gravitational waves~(GWs) from compact binary coalescences~\cite{LIGOScientific:2018mvr,LIGOScientific:2016lio,Cai:2017cbj,Bian:2021ini}. However, resolvable GW events only account for a small fraction of all GW signals reaching our detectors, and most are below confusion limit~\cite{2022Galax..10...34R}. The superposition of GWs from a large number of weak, independent, and unresolved sources can form the stochastic gravitational wave backgrounds~(SGWBs), which are the general prediction of various astrophysical processes and violent physical processes in the early Universe, such as first-order phase transitions~\cite{Kamionkowski:1993fg,Caprini:2015zlo,Hindmarsh:2015qta} and cosmic strings~\cite{Vachaspati:1984gt,Auclair:2019wcv,Damour:2000wa}. Detecting the relic GWs from the early Universe could enlighten new physical models at high energy scales~\cite{KAGRA:2021kbb,Maggiore:1999vm,Christensen:2018iqi}.
In Refs.~\cite{LIGOScientific:2017ikf,KAGRA:2021kbb,Romero:2021kby,LIGOScientific:2021nrg}, the corresponding constraints are given by LIGO-Virgo observations.
Besides, primordial curvature perturbations which seed the large-scale structure can also produce a SGWB through the scalar-tensor coupling at the second order~\cite{Ananda:2006af,Saito:2008jc,Baumann:2007zm,Kohri:2018awv,Domenech:2021ztg}.
In this paper, we use the latest LIGO-Virgo data to search for the scalar-induced SGWB originating from the nonattractor dynamics in the USR inflationary models.

Inflation is a successful framework of the very early Universe because it naturally provides the initial condition of the hot big-bang Universe, and the prediction of quantum fluctuations during inflation is also consistent well with the current observations of the CMB temperature anisotropies and large-scale structure~\cite{Lewis:1999bs,Bernardeau:2001qr}. Although the concept of inflation was proposed decades ago~\cite{Guth:1980zm,Linde:1981mu,Starobinsky:1980te}, the underlying physics model of inflation is far from being revealed. From the observation point of view, curvature perturbations are tightly constrained as $\sim 10^{-5}$ at the CMB scales~\cite{Planck:2018vyg}, while at small scales the constraints become much looser~\cite{Emami:2017fiy,Gow:2020bzo}. The observation of small-scale primordial perturbations is an indispensable measurement of the entire effective potential and dynamics of inflation.
The USR phase is a natural consequence of the non-attractor evolution of the inflaton field which can be realized in the scope of supergravity~\cite{Dalianis:2018frf,Gao:2018pvq,Wu:2021zta}, string theory~\cite{Cicoli:2018asa,Cicoli:2022sih,Ozsoy:2018flq}, modified gravity~\cite{Pi:2022zxs,Lin:2020goi,Yi:2022anu,Kawai:2021edk} and non-minimal couplings~\cite{Ezquiaga:2017fvi,Fu:2019ttf,Kawai:2022emp}, and also originate from small fluctuations in the inflationary potential, including inflection points~\cite{Choudhury:2013woa,Germani:2017bcs,Bhaumik:2019tvl,Bhaumik:2019tvl}, bumps~\cite{Mishra:2019pzq,Ozsoy:2020kat}, dips~\cite{Gu:2022pbo,Mishra:2019pzq} and steps~\cite{Kefala:2020xsx,Inomata:2021uqj,Cai:2021zsp,Inomata:2021tpx}. The USR regime generates a large $e$-folding number required to explain the horizon, flatness and monopole problems~\cite{Pattison:2018bct}.
	During the USR regime, the slow-roll approximation is violated and superhorizon curvature perturbations are exponentially amplified~\cite{Liu:2020oqe,Ozsoy:2019lyy,Byrnes:2018txb}, which is totally different from the nearly scale-invariant power spectrum predicted by the standard slow-roll inflationary model.
	Recently, the USR inflationary scenario has attracted much attention since amplified curvature perturbations lead to the formation of primordial black holes which
could potentially constitute part of or all dark matter~\cite{Carr:2016drx,Bird:2016dcv,Di:2017ndc,Garcia-Bellido:2017mdw,Hertzberg:2017dkh,Passaglia:2018ixg,Cai:2018dig,Fu:2020lob,Cai:2022erk,Figueroa:2021zah,Figueroa:2020jkf,Pi:2021dft,Pi:2021dft,Wang:2021kbh,Xu:2019bdp,Cheong:2019vzl,Braglia:2020eai,Escriva:2022duf}. Refs.~\cite{Wu:2021gtd,Balaji:2022rsy,Kristiano:2022maq,Franciolini:2022pav,Franciolini:2022tfm} find USR inflation can also achieve a successful baryogenesis.
	High-frequency GW observations provide a good chance to probe the late-time dynamics of inflation through scalar-induced GWs.
	The main purpose of this work is to place constraints on the inflationary potential of the USR phase and amplified curvature perturbations using the third LIGO-Virgo observing run data~(O3).
	
	For convenience, we choose $c=8\pi G=1$ throughout this paper.
	
	\section{Curvature perturbations and the SGWB from USR inflation}
	It is well-known that in slow-roll inflation, the Fourier modes of curvature perturbations, $R_{k}$, remain constant at superhorizon scales. However, in Ref.~\cite{Liu:2020oqe} we find that the time derivative of curvature perturbations, $\dot{R}_{k}$, is amplified exponentially during the USR phase at superhorizon scales.
The amplified power spectrum $P_{R}(k)$ peaks roughly at the mode which leaves the Hubble horizon at the beginning of the USR phase. The infrared side of the peak is asymptotically approximated as $P_{R}\propto k^{4}$~\cite{Liu:2020oqe,Ozsoy:2019lyy,Byrnes:2018txb}. This feature could distinguish USR inflation from other models. The ultraviolet side also has a power-law form, $P_{R}\propto k^{\beta}$, where $\beta$ depends on the inflationary potential. Let the USR regime ends at $\phi=\phi_{e}$, and the Taylor expansion of the inflationary potential around $\phi_{e}$ reads  $V(\phi)=b_0+b_1\left(\phi-\phi_e\right)+b_2\left(\phi-\phi_e\right)^2+\cdots$. Then, the spectral index $\beta$ can be expressed in terms of the expansion coefficients $\beta=3-\sqrt{9-24b_{2}/b_{0}}$, which is nearly constant since the inflaton rolls very slowly during and soon after the USR regime.
	
	The amplification of small scale $P_{R}(k)$ predicts a detectable strong SGWB.
	In the Newton gauge, the perturbed metric reads
	\begin{equation}
		\begin{split}
			d s^{2}= & a^{2}(\tau)\left\{-(1+2 \Phi) d \tau^{2} \right.\\
			&\left.+\left[(1-2 \Phi) \delta_{i j}+\frac{1}{2} h_{i j}\right] d x^i d x^j\right\}\,.
		\end{split}
	\end{equation}
Here, we neglect tensor perturbations from quantum fluctuations during inflation which is far from being observed in the LIGO band. In the scenario of USR inflation, scalar perturbations are immensely amplified so that GWs sourced by scalar perturbations are much stronger than that from quantum fluctuations.
	Tensor perturbations are coupled to scalar perturbations at the second order of the Einstein equations.
	For each of the two polarization patterns ($+$ and $\times$), the equation of motion of the Fourier modes of $h_{ij}$ reads
	\begin{equation}
	\label{eq:2}
		h_{\boldsymbol{k}}^{\prime \prime}+2 \mathcal{H} h_{\boldsymbol{k}}^{\prime}+k^2 h_{\boldsymbol{k}}=2 \mathcal{P}_{i j}^{l m} e^{i j} T_{l m}(\boldsymbol{k}, \tau)\,,
	\end{equation}
	where a prime denotes the derivative with respect to the conformal time $\tau$, $\mathcal{P}_{i j}^{l m}$ is the projection operator to the transverse-traceless part,
and the source term $T_{lm}$ comes from the second-order perturbation of the Einstein equation
	\begin{equation}
	\label{eq:3}
		T_{l m}=-2 \Phi \partial_l \partial_m \Phi+\partial_l\left(\Phi+\mathcal{H}^{-1} \Phi^{\prime}\right) \partial_m\left(\Phi+\mathcal{H}^{-1} \Phi^{\prime}\right)\,.
	\end{equation}
The equation of motion for $\Phi$ in the radiation-dominated era reads
	\begin{equation}
	\label{eq:4}
		\Phi_{\mathbf{k}}^{\prime \prime}(\eta)+\frac{4}{\eta} \Phi_{\mathbf{k}}^{\prime}(\eta)+\frac{1}{3} k^2 \Phi_{\mathbf{k}}(\eta)=0\,,
	\end{equation}
	where we have neglected entropy perturbations.
	
	Using the method proposed in Ref.~\cite{Kohri:2018awv}, we numerically obtain the energy spectrum of scalar-induced GWs,
	$\Omega_{\mathrm{GW}}\equiv \frac{1}{\rho_{c}}\frac{d\rho_{\mathrm{GW}}}{d\ln f}$, where $\rho_{c}$ is the critical energy density of the Universe and $\rho_{\mathrm{GW}}=\langle h_{ij,k}h_{ij,k}\rangle/(16a^{2})$ is the energy density of GWs. Here $\langle...\rangle$ denotes the 
spatial and oscillation average.
	
	\section{Data Analysis}
	We adopt the Markov chain Monte Carlo (MCMC) method to sample the parameter space and use the Bayesian approach to estimate the model parameters. The likelihood function is given by
	\begin{eqnarray}
		&&\nonumber p(\hat{C}^{IJ}(f)|\bf{\Theta}, \lambda)\propto \\
		&&\exp\left[-\frac{1}{2}  \sum_{IJ}\sum_{f}\left(\frac{\hat{C}^{IJ}(f)-\lambda \Omega_{\mathrm{GW}}(f, \bf{\Theta)}}{\sigma_{IJ}(f)}\right)^2\right]\,,
	\end{eqnarray}	
where $\hat{C}^{IJ}(f)$ denotes the cross-correlation statistic from the data for the baseline $IJ$ and $\sigma^2_{IJ}(f)$ is the variance of $\hat{C}^{IJ}(f)$.
We use the $\hat{C}^{IJ}(f)$ data from LIGO-Virgo O3 run, which includes three baselines: HL, HV and LV.
The $\hat{C}^{IJ}(f)$ data is model-independent and public available at \cite{dcc}. 
Here $\Omega_{\mathrm{GW}}(f, \bf{\Theta})$  describes the model for the SGWB with the parameter set $\bf{\Theta}$~\cite{KAGRA:2021kbb}. $\lambda$ characterizes the calibration uncertainties.
In our analysis we assume $\lambda$ is Gaussian distributed, and then it can be marginalized analytically~\cite{Whelan:2012ur}.
	
	Instead of numerically calculating the precise $P_{R}(k)$ from the inflationary potentials,
in our analysis we adopt the analytic approximation of $P_{R}(k)$ in the USR inflationary models obtained in Ref.~\cite{Liu:2020oqe} to reduce the computational cost of MCMC.
In this case $P_{R}(k)$ can be parameterized with the following form
	\begin{equation}
		\label{eq:param}
		P_R(k)=P_{R\mathrm{p}}\cfrac{(\alpha+\beta)^\gamma}{\left[\beta (k/k_{\mathrm{p}})^{-\alpha/\gamma}+\alpha (k/k_{\mathrm{p}})^{\beta/\gamma}\right]^\gamma}\,,
	\end{equation}
	where $\alpha$ and $\beta$ are the asymptotic spectral indexes for $k<k_{\mathrm{p}}$ and $k>k_{\mathrm{p}}$, respectively.
Note $P_R(k)$ reaches the peak value $P_{R\mathrm{p}}$ at $k=k_{\mathrm{p}}$.
USR inflation predicts that the infrared spectral index $\alpha=4$.
The parameter $\gamma$ controls the smoothness of $P_{R}(k)$ around the peak.
We find that Eq.~(\ref{eq:param}) with $\gamma=2.6$ can fit well the numerical results of $P_{R}(k)$
and hence we fix $\gamma$ in our anlaysis.
Therefore, the remaining free parameters are $P_{R\mathrm{p}}$, $k_{\mathrm{p}}$ and $\beta$,
which characterize the $e$-folding number, the beginning time and the potential shape of USR, respectively.
Note that the expression~\eqref{eq:param} captures the universal signature of USR inflation and is independent of specific models.
Although the approximation~\eqref{eq:param} deviates from the precise numerical result of $P_{R}(k)$ around the peak,
our analysis shows that such a slight deviation has a negligible effect on the energy spectrum of GWs.

Given the power spectrum of curvature perturbations $P_R(k)$,
in principle, we can numerically calculate $\Omega_{\mathrm{GW}}(f, \bf{\Theta})$ by the integration with respect to the wave numbers.
However, we need to repeatedly calculate the integration to obtain the values of $\Omega_{\mathrm{GW}}(f, \bf{\Theta})$
in the frequency bounds considered in our analysis.
This takes too much time to perform one MCMC step.
Hence we just compute the peak value of $\Omega_{\mathrm{GW}}(f, \bf{\Theta})$, i.e., $\Omega_{\mathrm{p}}$,
and approximate $\Omega_{\mathrm{GW}}(f, \bf{\Theta})$ similar to the parameterization~(\ref{eq:param}).
Except for the peak value, other parameters characterizing the curve of $\Omega_{\mathrm{GW}}(f, \bf{\Theta})$ are directly determined by $P_R(k)$.
The peak frequency is given by $f_{\mathrm{p}} \approx 0.03\,\mathrm{Hz}\cfrac{k_{\mathrm{p}}}{2\times 10^7\,\mathrm{pc}^{-1}}$
and the UV spectral index of $\Omega_{\mathrm{GW}}(f, \bf{\Theta})$ is given by $\beta' = 2\beta$.
In fact, the current data is not very sensitive to the exact form of $\Omega_{\mathrm{GW}}(f, \bf{\Theta})$.
We have checked that this approximation captures the key features of $\Omega_{\mathrm{GW}}(f, \bf{\Theta})$
and has a negligible effect on our constraint results.

The priors adopted in our analysis are summarized in Table \ref{table:prior}. The peak value of the power spectrum of curvature perturbations, $P_{R\mathrm{p}}$, should be smaller than $\mathcal{O}(1)$ by definition. From the theoretical point of view, we have no prior information about the location of the peak frequency $f_{\mathrm{p}}$.
Therefore, in this work, we investigate the following two cases.
One is that $f_{\mathrm{p}}$ falls in the LIGO-Virgo sensitive band,
which will impose strong constraints on the model parameters.
The other is to vary $f_{\mathrm{p}}$ by 6 orders of magnitude to explore a larger parameter space.
Obviously, constraints become weak as $f_{\mathrm{p}}$ is far from the LIGO-Virgo sensitive band.
Besides, $\beta$ varies from 0 to 10, the value of $\alpha$ is fixed to be $4$ by theoretical prediction,
and the value of $\gamma$ is fixed to be $2.6$ which make the parameterization~\eqref{eq:param} best fit the numerical result.
	
	\begin{table}
		\begin{center}
			\caption{\label{table:prior}Priors on the model parameters. For physical consideration, we restrict $P_{R\mathrm{p}}$ to be smaller than ${\cal{O}}(1)$. We use two priors for $f_{\mathrm{p}}$, one is uniformly distributed over the LIGO-Virgo most sensitive band, and the other is log-uniformly distributed across 6 orders of magnitude to scan a larger parameter space.
}
			\begin{tabular}{cc}
				\hline \hline
				Parameters& Priors \\
				\hline
				$P_{R\mathrm{p}}$ & Log-Uniform($10^{-6},1$) \\
				$f_{\mathrm{p}}$ & Log-Uniform($10^{-2},10^4\,\mathrm{Hz}$),Uniform($0,256\,\mathrm{Hz}$) \\
				$\alpha$ & 4\\
				$\beta$ & Uniform(0,10) \\
				$\gamma$ & 2.6\\
				\hline \hline
			\end{tabular}
		\end{center}
	\end{table}

	\section{Results}
\begin{figure}[h]
		\includegraphics[width=0.45\textwidth]{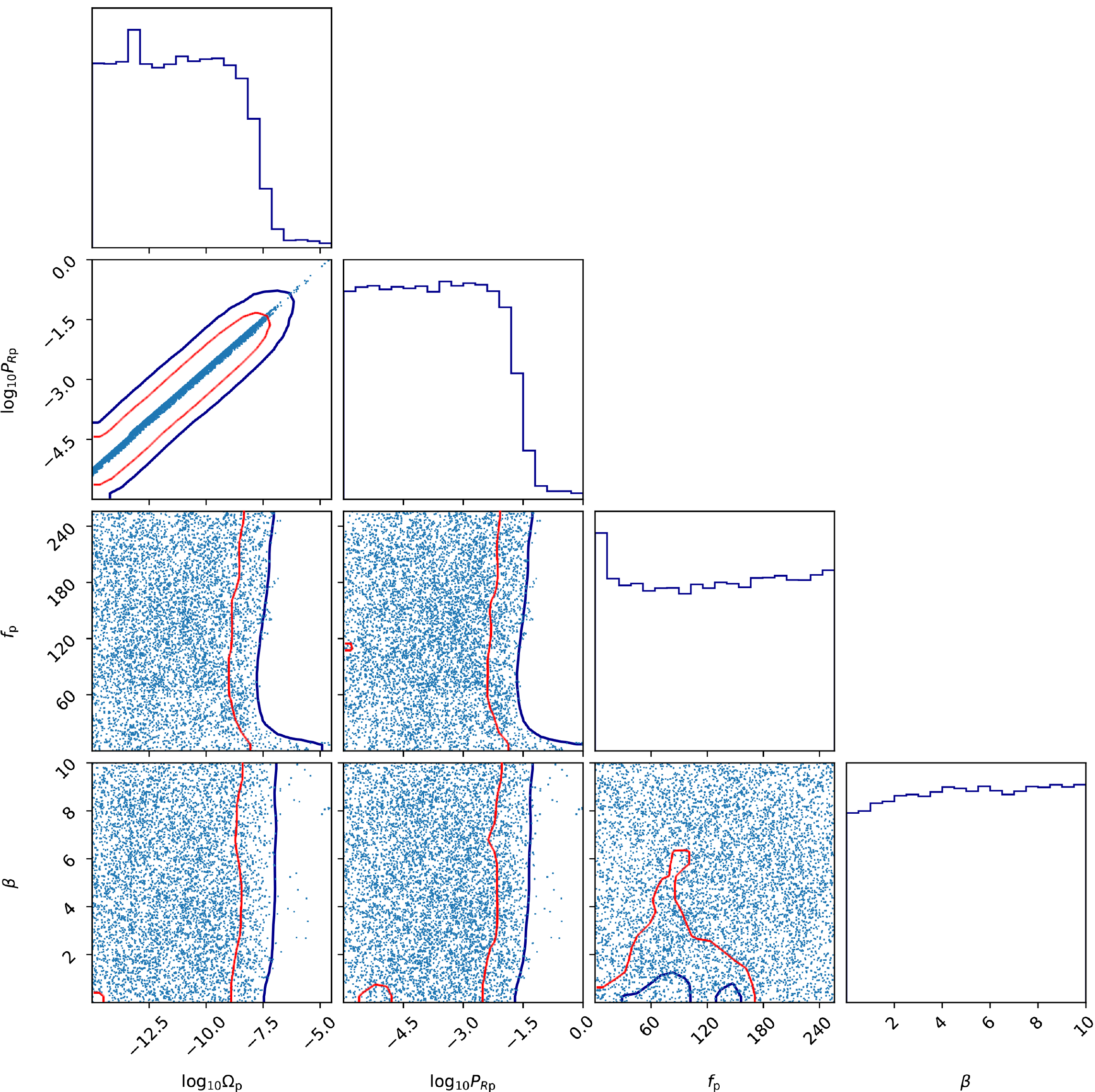}
		\caption{Posterior distribution of three input parameters and $\Omega_{\mathrm{p}}$.
 The peak frequency $f_{\mathrm{p}}$ is uniformly sampled. The red and blue lines denote the $68\%$ and $95\%$ confidence regions, respectively.}
		\label{fig:uniform}
	\end{figure}

\begin{figure}[h]
		\includegraphics[width=0.45\textwidth]{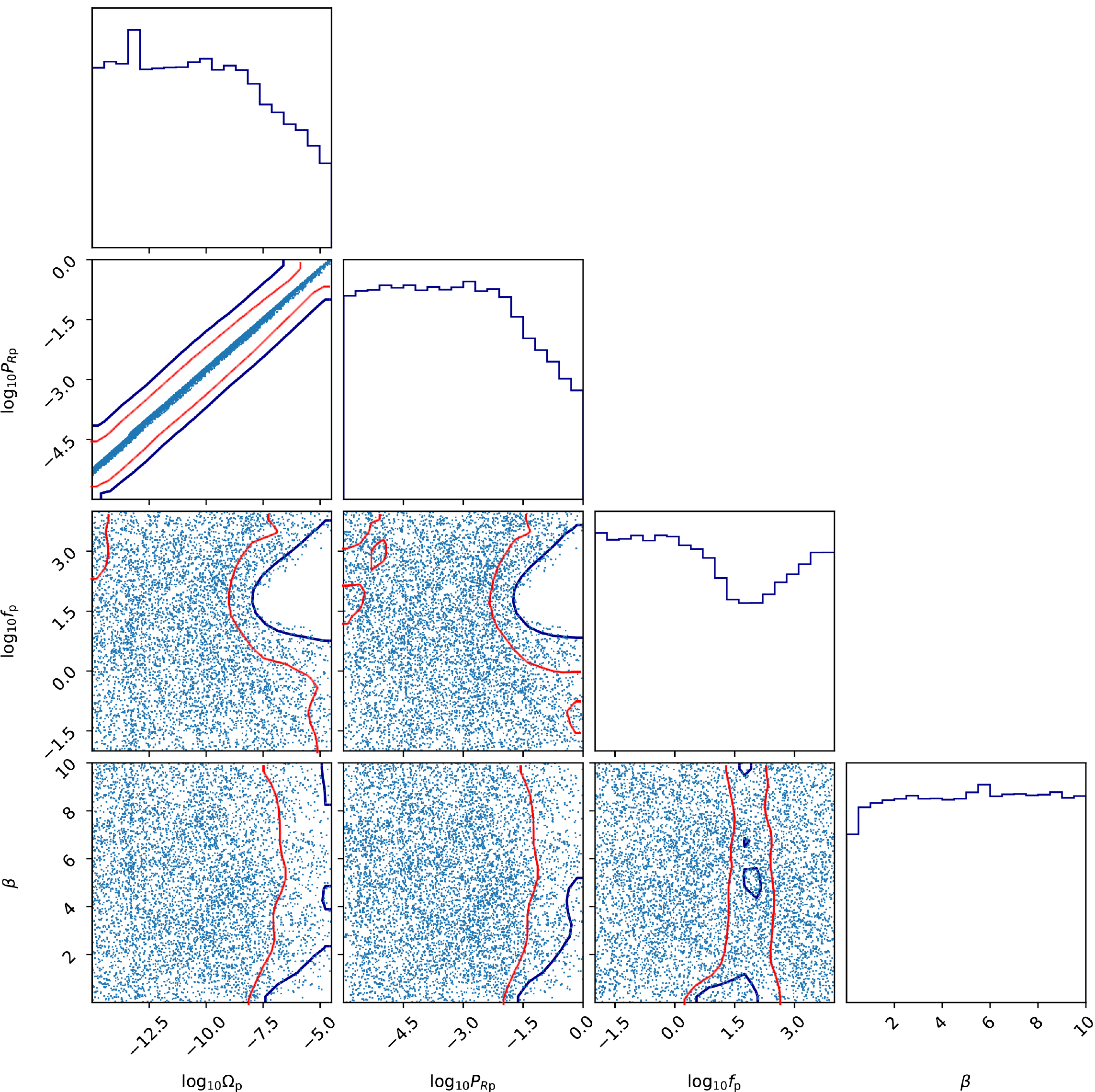}
		\caption{Posterior distribution of three input parameters and $\Omega_{\mathrm{p}}$.
 The peak frequency $f_{\mathrm{p}}$ is log-uniformly sampled. The red and blue lines denote the $68\%$ and $95\%$ confidence regions, respectively.}
		\label{fig:log}
	\end{figure}

	We use GW data from the third LIGO-Virgo observation run to constrain the model parameters listed in Table~\ref{table:prior} and the corresponding physical parameters of USR inflation.
	The analytical calculation for $P_{R}(k)$ predicted in USR inflation
implies that the amplification rate of $P_{R}$ at the peak is $e^{6\Delta N}$,
where $\Delta N$ denotes the $e$-folding number of the USR regime.
We utilize LIGO-Virgo observational data to give constraints on $\Delta N$
in terms of $\beta$ and the $e$-folding number $N$ of the end of the USR regime.
	The 1-dimensional marginalized posterior distribution and 2-dimensional contours are shown in Fig.~\ref{fig:uniform} and Fig.~\ref{fig:log} for the uniform and log-uniform sampling of $f_{\mathrm{p}}$.
The red and blue lines represent the $1\sigma$ and $2\sigma$ confidence regions, respectively.
The white regions are ruled out at the $95\%$ confidence level.

For uniformly sampled $f_{\mathrm{p}}$  between $0-256 ~\mathrm{Hz}$,
since $f_{\mathrm{p}}$ is within the most sensitive band of LIGO-Virgo,
constraints on $\Omega_{\mathrm{p}}$ and $P_{R\mathrm{p}}$ are strong and insensitive to $f_{\mathrm{p}}$ and $\beta$.
However, the 1-dimensional posteriors of $f_{\mathrm{p}}$ and $\beta$ are almost flat,
which indicates that current data gives weak constraints on these two parameters.
This is because in this case the prior of $f_{\mathrm{p}}$ is close to the sensitivity bands of LIGO and Virgo and no sign of any SGWB has been observed.

If $f_{\mathrm{p}}$ is log-uniformly sampled,
constraints on $\Omega_{\mathrm{p}}$ and $P_{R\mathrm{p}}$ are sensitive to $f_{\mathrm{p}}$ and $\beta$.
When the peak location $f_{\mathrm{p}}$ starts to move away from the LIGO-Virgo sensitivity band, the GW energy density in that band becomes much smaller than the peak $\Omega_{\mathrm{p}}$, especially for large $\beta$.
Then, the constraints on $\Omega_{\mathrm{p}}$ and $P_{R\mathrm{p}}$ become weak.
Likewise, the trough in the 1-dimensional posterior of $f_{\mathrm{p}}$ cannot be interpreted as an exclusion region,
which is also because LIGO-Virgo gives the strongest constraint in its sensitive band.

We present the upper limits at the $95\%$ confidence level for the parameters $\Omega_{\mathrm{p}}$ and $P_{R\mathrm{p}}$ in Table~\ref{table:constraint1}. In the case of log-uniformly sampling,
the upper limits are obtained by marginalizing other parameters,
and as mentioned before, the constraints on $\Omega_{\mathrm{p}}$ and $P_{R\mathrm{p}}$ depend on other parameters in this case. So one should refer to the contours in Fig.~\ref{fig:log} for more accurate constraints.
In the case of uniformly sampling, constraints on $\Omega_{\mathrm{p}}$ and $P_{R\mathrm{p}}$ do not change when marginalizing over $\beta$ and $f_p$,
so we take this as our main result.
	
\begin{figure}[h]
		\includegraphics[width=0.45\textwidth]{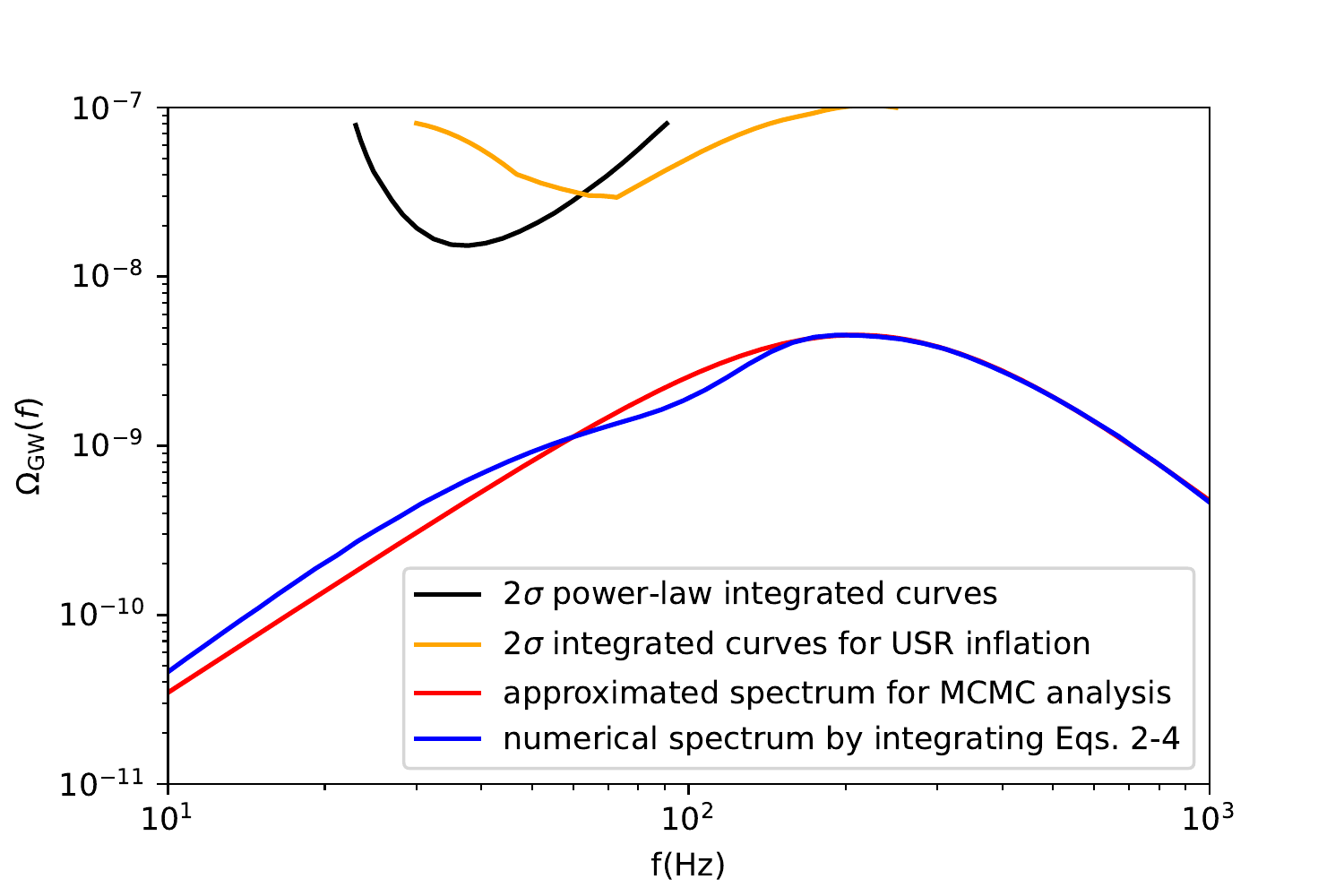}
		\caption{
Exact energy spectrum (blue) of scalar-induced GWs by numerically integrating Eqs.~(\ref{eq:2}) -  (\ref{eq:4}) and the approximate energy spectrum (red) for MCMC analysis with the same parameters. The orange and black curves represent the LIGO-Virgo constraint curves for USR inflation and the power-law model, respectively.}
		\label{fig:spectrum}
	\end{figure}

We show the energy spectra of scalar-induced GWs and the LIGO-Virgo constraint curves in Fig.~\ref{fig:spectrum}.
The blue curve is obtained by numerically integrating Eqs.~(\ref{eq:2}) -  (\ref{eq:4}),
and the red one is the corresponding approximate spectrum used for MCMC analysis.
For both blue and red curves, the model parameters are set the same as Fig.~3 of \cite{Liu:2020oqe}.
The approximate spectrum captures the key features of $\Omega_{\mathrm{GW}}(f, \bf{\Theta})$
and the difference between these two curves is small.
Besides, by using the method proposed in
\cite{Thrane:2013oya}, we obtain the 2$\sigma$ integrated curve, which is the upper envelop of GW energy spectra with the parameters set to the 2$\sigma$ upper limit.
The orange curve denotes the integrated curve for USR inflation (for the case that $f_{\mathrm{p}}$ is uniformly sampled)
and the black one denotes the integrated curve for
the power-law model (the solid black line in Fig.~5 of \cite{KAGRA:2021kbb}).
Since the integrated curve depends on the shape of the spectrum in the analysis band,
the orange and black curves are different but in the
same order of magnitude.

\begin{figure}[h]
		\includegraphics[width=0.45\textwidth]{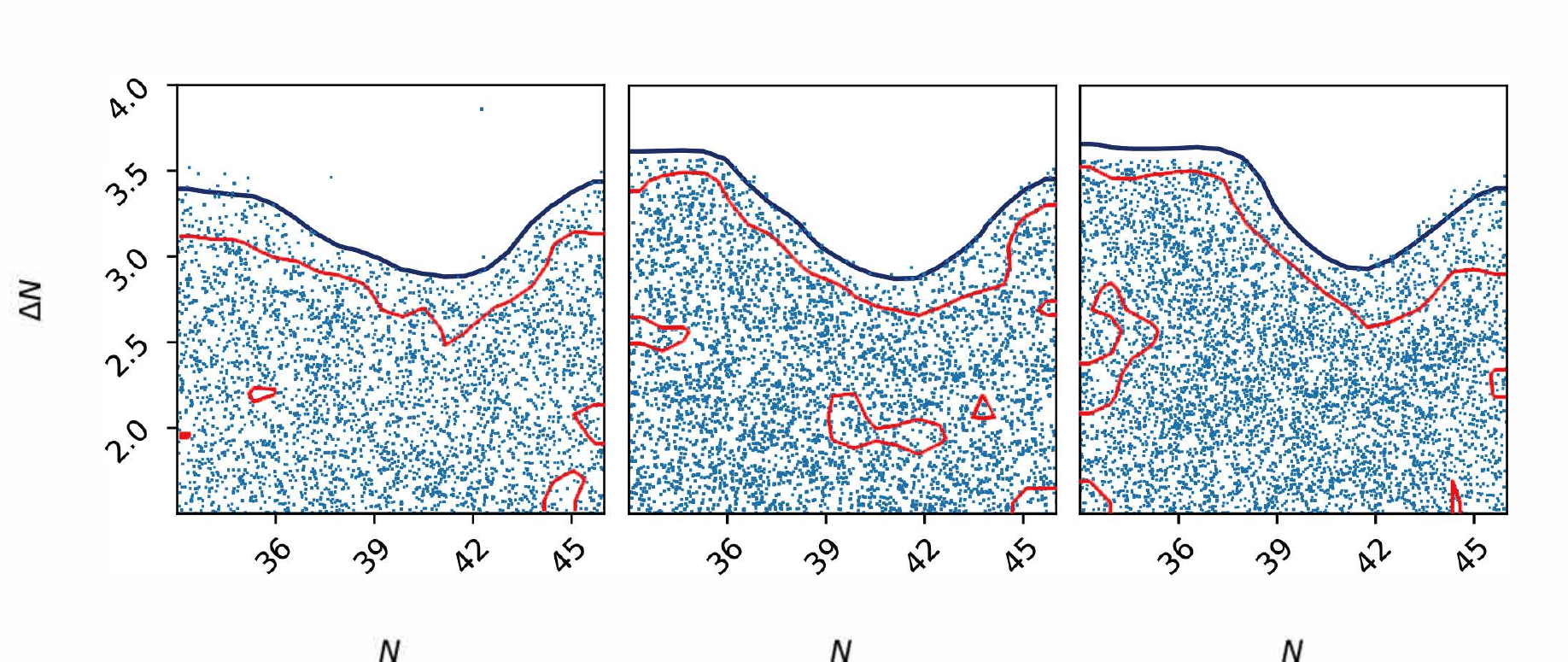}
		\caption{Posterior distribution of $N$ and $\Delta N$ for the log-uniformly sampling of $f_{\mathrm{p}}$.
The red and blue lines denote the $68\%$ and $95\%$ confidence regions, respectively. Here we fix $\beta= 0.5, 1, 2$ from left to right.}
		\label{fig:N_Delta_N}
	\end{figure}

Additionally, we give constraints on the physical parameters of USR inflation, $N$ and $\Delta N$, as shown in Fig.~\ref{fig:N_Delta_N}.
Since the uniformly and log-uniformly sampling give similar results in the overlapping parameter space,
we only consider the log-uniformly sampling to show constraints in the large parameter space.
We choose three values of $\beta$ to illustrate how the constraints depend on $\beta$.
We give the best upper limits on $\Delta N$ when $N=41.6,41.7,41.1$, respectively,
in the cases of $\beta=0.5,1,2$, as shown in Fig.~\ref{fig:N_Delta_N}.
The upper limits on $\Delta N$ for the strongest constraint are listed in Table~\ref{table:constraint2}.
For smaller $\beta$, $P_{R}(k)$ decreases more slowly with $k$ in the ultraviolet region,
which results in a stronger SGWB at the scales $k\gtrsim k_{\mathrm{p}}$.
Thus, for smaller $\beta$, the constraints on $\Delta N$ become stronger in a larger range of $N$.

	\begin{table}
		\begin{center}
			\caption{\label{table:constraint1}Upper limits on the parameters $\Omega_{\mathrm{p}}$ and $P_{R\mathrm{p}}$ at the $95\%$ confidence level for the uniformly and log-uniformly sampled cases.}
			\begin{tabular}{ccc}
				\hline\hline
				~~~& ~~~Uniform Prior ~~~ & ~~~  Log-Uniform Prior ~~~\\
				\hline
				$\log_{10}\Omega_{\mathrm{p}}$ & $-7.8 $    &  $-5.4 $       \\
				$\log_{10}P_{R\mathrm{p}}$  & $-1.7$    &  $-0.49$       \\
				\hline \hline
			\end{tabular}
		\end{center}
	\end{table}
	
	\begin{table}
		\begin{center}
			\caption{\label{table:constraint2}Upper limits on $\Delta N$ at the 95\% confidence level for the log-uniformly sampled case.
We give the best upper limits on $\Delta N$ when $N=41.6,41.7,41.1$, respectively,
in the cases of $\beta=0.5,1,2$, as shown in Fig.~\ref{fig:N_Delta_N}. }
			\begin{tabular}{ccc}
				\hline\hline
				~~~~~~~~~~$\beta$ ~~~~~~~~~~& ~~~~~~~~~~$N$ ~~~~~~~~~~ & ~~~~~~~~~~  $\Delta N$ ~~~~~~~~~~\\
				\hline
				$0.5$ & $41.6$    &  $2.90 $       \\
				$1$ & $41.7$    &  $2.87$       \\
				$2$ & $41.1$    &  $ 2.99$       \\
				\hline \hline
			\end{tabular}
		\end{center}
	\end{table}
	
	\section{Conclusions and discussions}
We search for the scalar-induced GW signals from USR inflation in LIGO-Virgo O3 data.
If we assume the peak frequency of GWs is within the band of $0 - 256 ~\mathrm{Hz}$,
we can obtain strong constraints on the $e$-folding number of the USR phase, $\Delta N \lesssim 2.9$ at the 95\% confidence level,
and the peak of the power spectrum of curvature perturbations amplified during the USR phase,
$\log_{10}P_{R\mathrm{p}}<-1.7$ at the scales $2.9\times10^5 ~\mathrm{pc^{-1}} \lesssim k \lesssim 1.7\times10^{11}~\mathrm{pc^{-1}}$.
We also search for the GW signals in the frequency bound of $10^{-2}-10^4$ Hz with the log-uniformly sampling of $f_{\mathrm{p}}$.

In our analysis we use data from the third LIGO-Virgo observing run.
Since the sensitivity of the detectors has been improved significantly after the first two observing runs,
including data from the first two observing runs does not change our main conclusions.
Specifically, the constraints would be more stringent by roughly $10\%$ after including O1 and O2 data, see Table 1 of \cite{KAGRA:2021kbb} as an example.

Pulsar timing arrays are sensitive to nanohertz GWs,
which complement LIGO-Virgo.
Recently the analysis of the NANOGrav 12.5-yr data set show
definite evidence for a common-spectrum stochastic signal across pulsars~\cite{NANOGrav:2020bcs}.
The 12.5-yr NANOGrav data allows us to explore a new region of parameter space of USR inflation.

Unlike the superhorizon amplification of curvature perturbations in USR inflation, subhorizon curvature perturbations may also be amplified by parametric resonance in either periodically oscillating sound speed~\cite{Cai:2018tuh,Cai:2019jah} or effective potential~\cite{Cai:2019bmk,Zhou:2020kkf}.
The profile of $P_{R}(k)$ may be a sharp peak or a broad plateau.
Constraints on $P_{R\mathrm{p}}$ with $\beta\gtrsim 0$ considered in this work apply to the plateau case
since the infrared GW spectrum in both cases is proportional to $k^{3}$.
However, for a $\delta$-function like $P_{R}(k)$, one may obtain stronger constraints on $P_{R\mathrm{p}}$ for $k>k_{\mathrm{p}}$ since $\Omega_{\mathrm{GW}}\propto k^{2}$ for $k\ll k_{\mathrm{p}}$ and GWs in the infrared regions become stronger~\cite{Cai:2019cdl}.

The energy spectrum of induced GWs quadratically depends on $P_{R}(k)$,  $P_{R}(k)$ amplified by USR inflation, we numerically obtain the approximate result $\Omega_{\mathrm{GW}}\approx0.5 (g_{0}/g_{*})^{1/3}\Omega_{r}P_{R}^{2}$, where $\Omega_{r}=9.2\times10^{-5}$ is the energy fraction of radiation at pressent, $g_{0}$ and $g_{*}$ are the relativistic degrees of freedom at pressent and the GW production time.
Thus, we can roughly estimate the ability to constrain USR inflation and $P_{R}(k)$ in the future GW detection projects, $P_{R\mathrm{p}}\lesssim 10^{-3.6}$ for LISA~\cite{Audley:2017drz} and Taiji~\cite{Guo:2018npi}, $P_{R\mathrm{p}}\lesssim 10^{-3.3}$ for Cosmic Explorer~\cite{Reitze:2019iox} and Einstein Telescope~\cite{Punturo:2010zz}, $P_{R\mathrm{p}}\lesssim 10^{-5.5}$ for DECIGO~\cite{Kawamura:2011zz} and Big Bang Observer~\cite{phinney2004big} and $P_{R\mathrm{p}}\lesssim 10^{-5.2}$ for Square Kilometre Array~\cite{Carilli:2004nx}.
	
	\acknowledgments
	This work is supported in part by the National Key Research and Development Program of China Grant No. 2020YFC2201501, in part by the National Natural Science Foundation of China under Grant No. 12075297, No. 12235019 and No. 12105060.

	\bibliographystyle{apsrev}
	\bibliography{ms}
\end{document}